\documentclass[%
 aip,
 amsmath,amssymb,
 reprint,%
]{revtex4-1}

\usepackage{graphicx}% Include figure files
\usepackage{dcolumn}% Align table columns on decimal point
\usepackage{bm}% bold math
\usepackage[utf8]{inputenc}
\usepackage[T1]{fontenc}
\usepackage{mathptmx}
\usepackage{etoolbox}

\usepackage{subcaption}
\usepackage{float}
\usepackage{color}
\usepackage{ulem}

\makeatletter
\def\@email#1#2{%
 \endgroup
 \patchcmd{\titleblock@produce}
  {\frontmatter@RRAPformat}
  {\frontmatter@RRAPformat{\produce@RRAP{*#1\href{mailto:#2}{#2}}}\frontmatter@RRAPformat}
  {}{}
}%
\makeatother
\begin{document}

\preprint{AIP/123-QED}

\title[]{Effect of the electronic pressure on the energy and magnetic moment of charged test particles in turbulent electromagnetic fields}
% Force line breaks with \\
\author{B. Balzarini}
 \affiliation{Departmento de F\'isica, Facultad de Ciencias Exactas y Naturales, Universidad de Buenos Aires, 1428 Buenos Aires, Argentina}%Lines break automatically or can be forced with \\
\author{F. Pugliese}%
\email{fpugliese@df.uba.ar}
\affiliation{Departmento de F\'isica, Facultad de Ciencias Exactas y Naturales, Universidad de Buenos Aires, 1428 Buenos Aires, Argentina}%Lines break automatically or can be forced with \\
\affiliation{Instituto de F\'isica del Plasma (INFIP), CONICET, 1428 Buenos Aires, Argentina%\\This line break forced with \textbackslash\textbackslash
}%

\author{P. Dmitruk}
\affiliation{Departmento de F\'isica, Facultad de Ciencias Exactas y Naturales, Universidad de Buenos Aires, 1428 Buenos Aires,  Argentina}%Lines break automatically or can be forced with \\
\affiliation{Instituto de F\'isica del Plasma (INFIP), CONICET, 1428 Buenos Aires, Argentina%\\This line break forced% with \\
}%

\date{\today}% It is always \today, today,

\begin{abstract}
In this work we perform direct numerical simulations of three-dimensional magnetohydrodynamics with a background magnetic field, representing solar wind plasma, and introduce test particles to explore how a turbulent electromagnetic environment affects them. Our focus is on the terms of the electric field present in the generalized Ohm's Law that are usually dismissed as unimportant. These are the Hall and the electronic pressure (EP) terms, but we concentrate primarily on the latter. We discover that the EP term generates an acceleration of the particles, which represent protons, in the direction parallel to the background magnetic field, in contrast to the known preferential perpendicular energization. By studying the electric field itself, we are able to detect the type of structures of the EP field that produce such parallel acceleration. These are thin and elongated structures placed on top of a monotonic and near-zero background. A statistical study to understand the real significance of the electronic pressure term is also performed.
\end{abstract}

\maketitle

\section{\label{sec:intro}Introduction}

Solar plasma and solar wind are astrophysical systems typically described macroscopically by means of the magnetohydrodynamic (MHD) model. It works for low frequency events since it addresses the system from the point of view of fluid dynamics. However, turbulent flows, such as the ones present in the sun, cover a huge range of spatio-temporal scales\cite{dissip, Matthaeus2021}. The MHD model is suitable for large scales, where the energy spectrum is consistent with the Kolmogorov homogeneous turbulence power law, but not for smaller scales.
Therefore, the macroscopic behavior of the plasma often requires the use of particle models to be adequately described. 
Kinetic models offer the most complete descriptions in such cases\cite{Banon2016, Kawazura2018, Nttil2022, Comisso2022}, but are computationally expensive and usually involve some level of approximation (for instance, not realistic ratio of ion-to-electron mass).
%Charged test particles are often used to obtain an approximation of the microscopic behavior of the plasma without the need to consider kinetic theory. And they can also represent groups of space particles passing through the solar plasma. Several studies have successfully adopted this approach to study the outer layers of the sun \cite{dmitruk2004, dmitruk2006_a, gonzales, gonzales2, lehe}.

\begin{figure}
    \centering
    \includegraphics[scale=0.55]{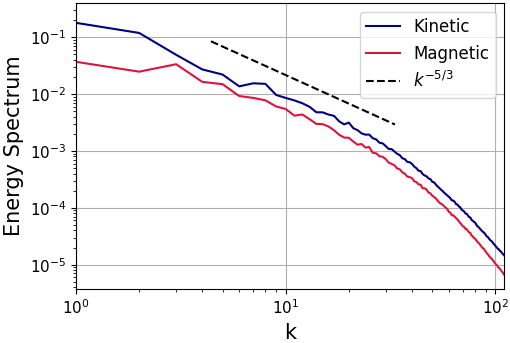}
    \caption{Kinetic and magnetic energy spectrum of the fluid when a stationary state is reached. The dotted line shows the Kolmogorov spectrum in the inertial range.}
    \label{fig:espectros}
\end{figure}

The test particle approach is computationally less demanding and useful for studying low density and energy particle populations, at least in comparison to the surrounding particles composing the bulk of the plasma.
Under such conditions, test particles can be used to obtain an approximation of the microscopic behavior of the plasma without the need to consider kinetic theory. 
An example of this are the solar energetic particles located inside the heliosphere, which are not likely to exhibit cooperative behavior or to be strongly scattered by wave-particle interactions\cite{Luhmann2003, Jarvinen2014}.
Several studies have successfully adopted this approach to study the outer layers of the sun \cite{dmitruk2004, dmitruk2006_a, gonzales, gonzales2, lehe}.
%Preliminary comparisons between kinetic and test particle models have been done in Gonzalez et al \textbf{CITAR CINETICO} for a simplified 2.5 dimensions and show good general agreement.

%\textcolor{black}{Solar energetic particles (SEPs) located inside the heliosphere have low density and energy. This is in comparison to surrounding particles which are considerably more powerful. SEPs are thus not likely to exhibit cooperative behavior or to be strongly scattered by wave-particle interactions\cite{Luhmann2003, Jarvinen2014}.}

It is known that there are turbulent magnetic fluctuations in these space plasmas and that they operate mainly in the directions perpendicular to the mean interplanetary magnetic field (background field), effect known as solar wind turbulence anisotropy \cite{mat3}. A consequence as well is a significantly higher energization of charged particles in the direction perpendicular to the background field \cite{dmitruk2004,pugliese2022}. Besides this, another phenomenon that occurs is the so-called coronal heating. This is the net energization of the coronal and solar wind plasma over time, and therefore, the progressive increase in its temperature \cite{parker, mat2}. Both previous effects are still cause for study and interest in solar plasma.

\begin{figure}
    \centering
    \begin{subfigure}[t]{0.46\textwidth}
        \centering
        \includegraphics[width=2.6in]{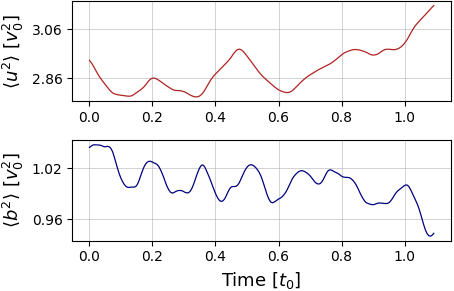}
        \caption{Simulation done without taking the Hall effect into account.}
        \vspace{4mm}
        \label{fig:estacionarioSinH}
    \end{subfigure}\hfill
    ~
    \begin{subfigure}[t]{0.46\textwidth}
        \centering
        \includegraphics[width=2.6in]{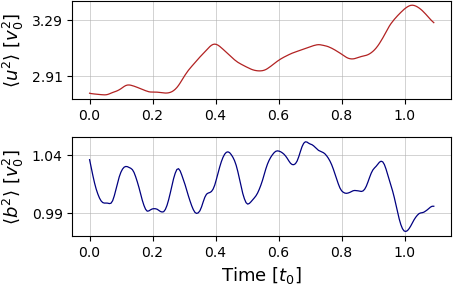}
        \caption{Simulation done taking the Hall effect into account.}
        \label{fig:estacionarioConH}
    \end{subfigure}
    \caption{Evolution of the kinetic and magnetic energy of the fluid during the time that test particles are present.}
    \label{averages_energies}
\end{figure}

In order to better understand this type of systems, in this work we perform direct three-dimensional numerical simulations based on pseudo-spectral methods \cite{parallel_pseudoespectral}.
Energy is injected into the system at the larger scales by mechanical and electromotive forces in order to drive turbulence. The natural mechanisms that promote such mechanical forces are, for instance, the footpoint motion at the photosphere\cite{Harvey1973, Dmitruk1997, Suzuki2006} and the fluid instabilities in the solar wind\cite{Goldstein1995, Marsch2006, Alexandrova2013}. Electromotive forces, on the other hand, may come from convective movements in the photosphere that affect the velocity field directly and the magnetic field indirectly through the frozen-in field, or large scale Alfvén waves in the solar wind\cite{Fermi1949, Hollweg1978}.
All parameters are chosen to be compatible with the solar wind plasma. The turbulent behaviour is not externally imposed, but instead, develops self-consistently and any coherent structure that arises is the result of the natural evolution of the plasma. To represent the interplanetary field, we will use a magnetic background field $\mathbf{B_0} = B_0\hat{z}$. We will have the weakly compressible MHD fields on the one hand (velocity, density, electric and magnetic fields) and the test particles, representing protons, on the other. The latter ones will be inserted into the system once it has reached a stationary turbulent state, and will be evolved as the fluid fields continue to develop.

We aim to understand the effect of a turbulent environment on test particles, in particular, on their energy and magnetic moment. As for the latter, it is the first adiabatic invariant, meaning that it stays approximately constant under slow spatio-temporal variations of the electromagnetic field\cite{RoedererZhang}. A turbulent flow does not fall under this category, and hence, the conservation of the magnetic moment is not to be expected \cite{dalena}. On the other hand, it has been demonstrated \cite{dmitruk2004} that particles gain energy at the dissipation scale through planar structures known as current sheets, formed along the background magnetic field direction. Moreover, compressibility was shown to play a role in particle energization, creating non-thermal particles \cite{lynn}. The interaction of the particles with coherent long lived structures has been found also to play an important role in the energization process \cite{Teaca2014,pugliese2022}. All of the above depicts some of the main processes by which charged particles can be affected by turbulent electromagnetic fields.

\begin{figure*}[htb]
    \centering
    \begin{subfigure}[t]{1\textwidth}
        \centering
        \includegraphics[width=5in]{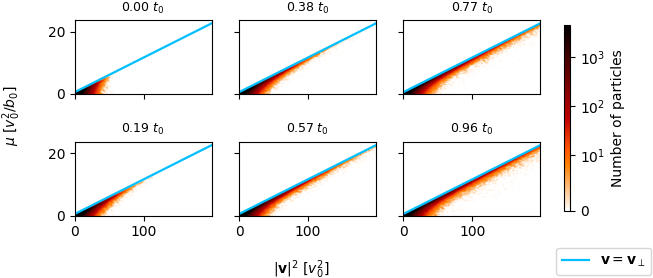}
        \caption{Simulation done without the Hall nor the electronic pressure terms.}
        \label{fig:KvsMu_SinPE}
    \end{subfigure}\hfill
    ~
    \begin{subfigure}[t]{1\textwidth}
        \centering
        \includegraphics[width=5in]{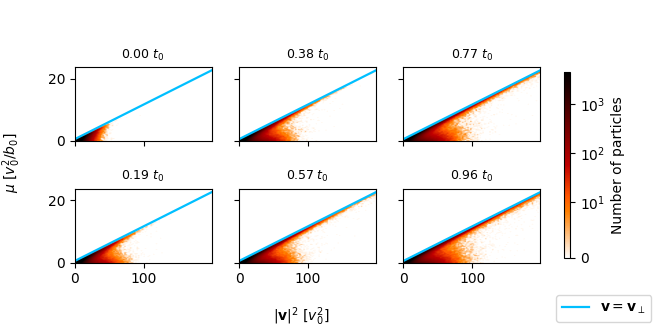}
        \caption{Simulation done with the electronic pressure term but not the Hall term.}
        \label{fig:KvsMu_ConPE}
    \end{subfigure}
    \caption{Two-dimensional histograms, at consecutive times, of the magnetic moment as a function of the energy for the $5\times 10^5$ test particles. The logarithmic color scale indicates the number of particles in each interval. The dotted line corresponds to a purely perpendicular velocity.}
    \label{averages energies}
\end{figure*}

Our main focus will be on how the test particles are affected by the different terms of the plasma's electric field. We will pay special attention to the Hall and electronic pressure terms, examining the role they play in particle acceleration. We are interested in how their effect differs in the directions parallel and perpendicular to the background magnetic field, given the anisotropy it introduces on the system.
We aim to better understand what is being ignored by dropping the Hall and EP terms, and how they contribute to the anisotropy of charged particle dynamics.
Our interest stems from the fact that these terms are often omitted in test particle studies.\cite{dmitruk2004, lehe, gonzales}
%\textcolor{black}{It is worth recalling that test particle studies often disregard both these terms.}

The present paper is organized as follows: Section \ref{models} introduces the models used to describe and simulate the plasma. All the equations and parameters used both for the fields (section \ref{fields}) and for the test particles (section \ref{t_particles}) are presented here. In section \ref{numerical results} the results of the numerical simulations are shown. We start by looking at the relationship between the magnetic moment and kinetic energy of the particles in \ref{relationship}. In \ref{average} we focus on the average energies of the particles and how they are influenced by the different terms of the electric field, while in section \ref{statistics} we turn our attention to the characteristics of the electric field itself. In \ref{test} we study the behaviour of individual test particles trying to reconcile this with the findings of the previous sections. Finally, section \ref{conclusiones} outlines the discoveries and conclusions of this work.

\section{\label{models}Models}

\subsection{\label{fields}The fields}

The description of the electromagnetic field is given by the compressible three-dimensional magnetohydrodynamic (MHD) equations, which is a one-fluid plasma model. It is composed of the equations \eqref{mhd1} to \eqref{mhd4}, which represent, respectively, the continuity equation, the equation of motion, the magnetic field induction equation and the equation of state. These describe the evolution of the plasma's density $\rho$, velocity field $\mathbf{u}$ and magnetic field $\mathbf{B}$. The total magnetic field is given by the magnetic fluctuations $\mathbf{b}$ plus a background field in the $z$-direction $\mathbf{B}_0$, meaning that $\mathbf{B} = \mathbf{B}_0 + \mathbf{b}$.

\begin{gather}
    \frac{\partial\rho}{\partial t} + \nabla\cdot (\rho \textbf{u}) = 0 \label{mhd1}\\
    \frac{\partial\textbf{u}}{\partial t}+(\mathbf{u}\cdot\nabla)\mathbf{u} = \frac{\textbf{J}\times\textbf{B}}{4\pi\rho} - \frac{\nabla p}{\rho} + \frac{\nu}{\rho}\bigg(\nabla^2\mathbf{u}+\frac{\nabla(\nabla\cdot\mathbf{u})}{3}\bigg) \textcolor{black}{+ \mathbf{f}} \label{mhd2}\\
    \frac{\partial\textbf{B}}{\partial t} = \nabla\times(\textbf{u}\times\textbf{B}) + \eta\nabla^2\textbf{B} - \frac{m_pc}{e}\nabla\times\left(\frac{\mathbf{J}\times\mathbf{B}}{4\pi\rho}\right) \textcolor{black}{+ \nabla\times \boldsymbol{F}} \label{mhd3}\\
    p\rho^{-\gamma} = \text{const.} \label{mhd4}
\end{gather}
In the previous expressions, $\mathbf{J} = \nabla\times\mathbf{B}$ represents the current density, $p$ is the pressure, $\nu$ the viscosity and $\eta$ the magnetic diffusivity.
\textcolor{black}{The large scale mechanical and electromotive forcing are $\mathbf{f}$ and $\nabla\times\boldsymbol{F}$, respectively.
We regard this electromotive forcing as arising from an externally induced electric field $\boldsymbol{F}$.}
As we consider the plasma ions to be mainly protons, their charge $e$ and mass $m_p$ appear in the Hall term of \eqref{mhd3}, along with the speed of light $c$.
Regarding the equation of state, we assume it to be adiabatic.
This means that $p\rho^{-\gamma} = p_0\rho_0^{-\gamma}$ with $\gamma = 5/3$ and $p_0$, $\rho_0$ a reference pressure and density respectively.

As mentioned in the introduction, we want to analyze the implications of either considering or ignoring the Hall and EP terms. Because of the barotropic hypothesis, the EP term can be written as a gradient, thus providing null contribution to the curl of the electric field. 
%And so including it or not does not change the fluid's equations, that is, equation \eqref{mhd3} is unaffected. 
Therefore, its inclusion in equation \eqref{mhd3} is not required.
As for the Hall effect, we simply choose to include or not the last term on the right-hand side of equation \eqref{mhd3} in each simulation.

\begin{figure}
    \centering
    \begin{subfigure}[t]{0.4\textwidth}
        \centering
        \includegraphics[width=2.7in]{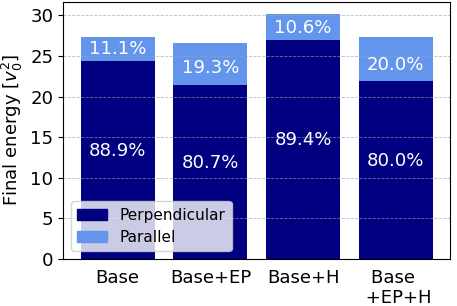}%width=2.82in
        \caption{Final value, in various cases, of the average energies (parallel and perpendicular to $\mathbf{B}_0$). In addition, the percentages of perpendicular and parallel energy with respect to the total energy are reported in each case.
        \vspace{4mm}}
        \label{fig:comparacion energias finales}
    \end{subfigure}\hfill
    ~
    \begin{subfigure}[t]{0.44\textwidth}
        \centering
        \includegraphics[width=3.27in]{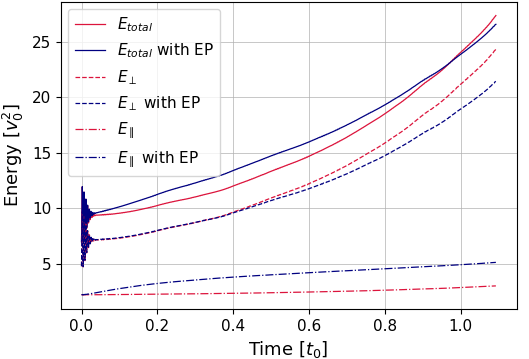}
        \caption{Evolution of the average energies (parallel and perpendicular to $\mathbf{B}_0$) in the Base and the EP case.}
        \label{fig:evolucion energias}
    \end{subfigure}
    \caption{Overview of the average energies over all particles.}
    \label{averages__energies}
\end{figure}

For the magnetic and velocity fields we use Alfvén speed units based on magnetic field fluctuations. These are defined as the rms (root mean squared) value $b_0 = \sqrt{\langle b^2\rangle_{\mathbf{x}, t}}$ and $v_0 = b_0/\sqrt{4\pi\rho_0}$, which is the Alfvén speed based on the 
rms magnetic field fluctuations. Here $\langle\bullet\rangle_{\mathbf{x}, t}$ is the space and time average during the whole simulation. Another characteristic speed of the problem is that of the Alfvén waves, that is, $v_A = B_0/\sqrt{4\pi\rho_0}$. We define the sonic Mach number $M_S = v_0/C_s$ where $C_s = \sqrt{\gamma p_0/\rho_0}$ is the speed of sound. To measure distances we take the isotropic MHD turbulence correlation length $L$, also known as energy containing scale, defined as $L = 2\pi\langle \int(E(k)/k)dk/\int E(k)dk \rangle_t$, where $E(k)$ is the energy spectral density at wavenumber $k$. Lastly, the unit we use to measure time is defined as the ratio between the unit length and the fluctuation Alfvén speed, so that $t_0 = L/v_0$.
Finally, the plasma has $\beta = (C_s/v_A)^2\approx 0.2$.
It is important to mention that these values vary only slightly between the simulations that consider the Hall effect and those that do not. This helps us compare the different results and we will take advantage of this by using the same set of units for all simulations.

%\textcolor{black}{It is important to mention that these values, which in turn define the units used throughout the work, vary only slightly between simulations that consider the Hall effect and those that do not. Let us also recall that the EP term does not affect the fluid's equations. Nevertheless, for consistency purposes, we will use the same set of units in every simulation.}

To solve the MHD equations numerically, we use a Fourier pseudospectral method, assuming periodic boundary conditions in a cube whose side is $L_{\text{box}}=2\pi$. This method guarantees exact energy conservation for the continuous time spatially discrete equations \cite{parallel_pseudoespectral}. For the discrete time integration a second-order Runge-Kutta method is used. In all the MHD simulations presented in this paper we considered a resolution of $N^3 = 512^3$ Fourier modes. In addition, the two-thirds rule truncation method is carried out to eliminate aliasing. In other words, the maximum wavenumber resolved is $\kappa = N/3$. By doing this we are able to reach values of kinematic Reynolds number, defined as  $R=v_0L\rho_0/\nu$, and magnetic Reynolds number, defined as $R_m=v_0L/\eta$, of $R=R_m=2370$. Defining the Kolmogorov dissipation wavenumber $k_d = (\epsilon_d(\rho_0/\nu)^3)^{1/4}$, where $\epsilon_d$ is the energy dissipation rate, the Reynolds numbers we use ensure resolution of the smallest scales (i.e., $\kappa>k_d$). Furthermore, the dissipation scale is defined as $l_d = 2\pi/k_d$.

\begin{figure*}
    \centering
    \begin{subfigure}[t]{0.5\textwidth}
        \centering
        \includegraphics[height=2.05in]{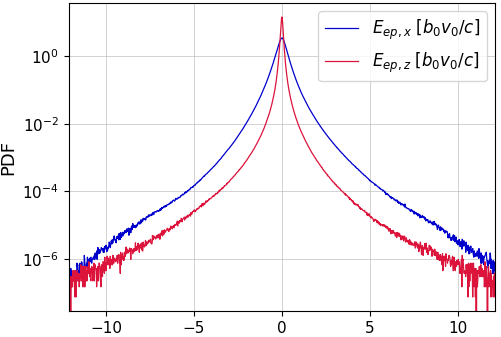}
        \caption{PDFs of the $E_{ep,x}$ and $E_{ep,z}$ fields.}
        \label{fig:Fep para y perp}
    \end{subfigure}%
    ~
    \begin{subfigure}[t]{0.5\textwidth}
        \centering
        \includegraphics[height=2.05in]{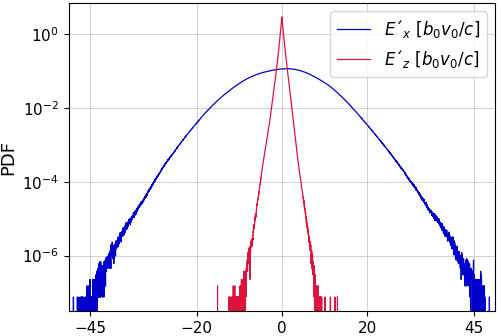}
        \caption{PDFs of the $E'_{x}$ and $E'_{z}$ fields.}
        \label{fig:E' para y perp}
    \end{subfigure}
    \caption{PDFs of the value of different fields at the position of the $500,000$ test particles, at all times. Simulation performed taking electronic pressure and Hall effect into account.}
    \label{histogramasss}
\end{figure*}

Regarding the MHD simulation's initial conditions, the system starts from null velocity and magnetic fluctuation fields $\mathbf{u}=\mathbf{b}=0$, and initial constant density $\rho = \rho_0$. To evolve from this initial state to a turbulent steady state, an external forcing is needed. We use a mechanical and an electromotive force, both generated with random phases in the Fourier $k$-shells $2\leq k\leq 3$ every correlation time $\tau$ \textcolor{black}{and thus each component can be written as}
\textcolor{black}{\begin{align}
 f_j &= \sum_{2\leq |\mathbf{k}| \leq 3 } f_0 \cos(\mathbf{k}\cdot\mathbf{x} + \phi_j(\mathbf{k})) \\
 F_j &= \sum_{2\leq |\mathbf{k}| \leq 3 } \frac{F_0}{|\mathbf{k}|} \cos(\mathbf{k}\cdot\mathbf{x} + \varphi_j(\mathbf{k})) \label{eq:em_forcing}
\end{align}}
\textcolor{black}{Here $f_0$ and $F_0$ are global amplitudes and the phases $\varphi_j(\mathbf{k})$ and $\phi_j(\mathbf{k})$ are randomly chosen for each component $j$ and wavenumber $\mathbf{k}$ independently.
The factor $1/|\mathbf{k}|$ in \eqref{eq:em_forcing} is included in order to prevent the Fourier coefficients of $\nabla\times\boldsymbol{F}$ from scaling as $|\mathbf{k}|$.}
We selected a value of $\tau\approx t_0/18$ that lays between the large eddy turnover time $t_0$ and the characteristic proton gyroperiod $\tau_p\approx t_0/200$ (see below). Intermediate forcing in time is achieved through linear interpolation between the forcing at the previous and at the next correlation time. 
By the time a stationary turbulent state is reached, \textcolor{black}{a new simulation with a higher background magnetic field $B_0$ is performed using this state as the initial condition, adjusting the forcing amplitudes to this new setup.}
The process is repeated until a \textcolor{black}{simulation with} $B_0/\langle b^2\rangle_{\mathbf{x},t} ^{1/2} = 9$ is achieved \textcolor{black}{and this is the final stationary state that will be used for our study}. Once we get to this state, $B_0/b_0 = 9$, $L_{\text{box}}/L\approx 2.55$, $\langle u^2\rangle_{\mathbf{x},t} ^{1/2}/v_0 \approx  1.67$ and $l_d/L\approx 1/60$. The energy spectrum of the fluid at this point can be seen in figure \ref{fig:espectros}.  As expected, it is found to be compatible with the Kolmogorov spectrum in the inertial range. 

\subsection{\label{t_particles}Test Particles}

Once the stationary turbulent state is reached, $5\times 10^5$ charged test particles are inserted in the fluid representing protons. They are evolved along with the plasma fields, yet their presence does not affect the fluid itself. The latter property is what defines the particles as test particles. During their evolution, the guide field $B_0$ is kept constant. To describe their dynamic within an electromagnetic field we use the non-relativistic Lorentz force:

\begin{gather}
    \frac{d\mathbf{v}}{dt} = \frac{q}{m}(\mathbf{E}+\mathbf{v}\times\mathbf{B}).
    \label{eq motion parts}
\end{gather}
Here, $\mathbf{v}$ is the velocity of the particle, $q$ is its charge and $m$ its mass. The magnetic field is obtained directly from the MHD equations. The electric field can be derived from the generalized Ohm's law. If we write it in a dimensionless form using a characteristic electric field defined as $E_0 = v_0b_0/c$, we obtain the following expression,

\begin{gather}
    \mathbf{E} = \frac{\mathbf{J}}{R_m} - \mathbf{u}\times\mathbf{B} + \frac{\epsilon}{\rho}\Big(\mathbf{J}\times\mathbf{B} - \frac{1}{\gamma M_S^2}\nabla p_e\Big). \label{ohm generalizado}
\end{gather}
The term proportional to $\mathbf{J}\times\mathbf{B}$ is known as the Hall term and the term proportional to $\nabla p_e$ is the electronic pressure (EP) term. 
%\textcolor{black}{For consistency, in simulations where the Hall effect is neglected for the particles, it will also be disregarded for the fluid, that is, in equation \eqref{mhd3}.} 
For consistency, in simulations where the Hall effect is neglected, it will be disregarded both in \eqref{mhd3} and \eqref{ohm generalizado}. 
The quantity $\epsilon = d_{i}/L$ stands for the Hall scale, where $d_{i} = m_p c/\sqrt{4\pi\rho_0e^2}$ is the proton inertial length. 
%In this last expression, $m_p$ and $e$ are the mass and charge of the proton, respectively. 
We will take $d_{i}$ to be equal to the dissipation scale. This assumption is supported by the fact that solar wind observations show $d_{i}\sim l_d$ \cite{dissip}. If we now define $p$ as the total fluid pressure, $p_e$ as the electronic pressure and $p_i$ as the ionic one, then $p=p_i+p_e$. In addition, we take the electrons and ions to be in thermal equilibrium, meaning that $p_i = p_e$.

\begin{figure}
    \centering
    \includegraphics[scale=0.65]{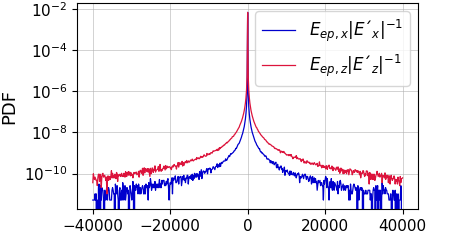}
    \caption{PDFs of the $E_{pe,x}/E'_x$ and $E_{pe,z}/E'_z$ fields at the position of the $500,000$ test particles, at all times. Simulation performed taking electronic pressure and Hall effect into account.}
    \label{fig: Fpe sobre Eprima para y perp}
\end{figure}

\begin{figure*}
    \centering
    \begin{subfigure}[t]{0.5\textwidth}
        \centering
        \includegraphics[height=2.35in]{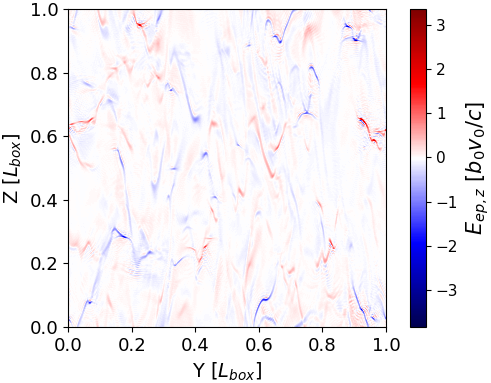}
        \caption{Cut in the $ZY$ plane for a given $X$ value.}
        \label{fetavertical}
    \end{subfigure}%
    ~ 
    \begin{subfigure}[t]{0.5\textwidth}
        \centering
        \includegraphics[height=2.35in]{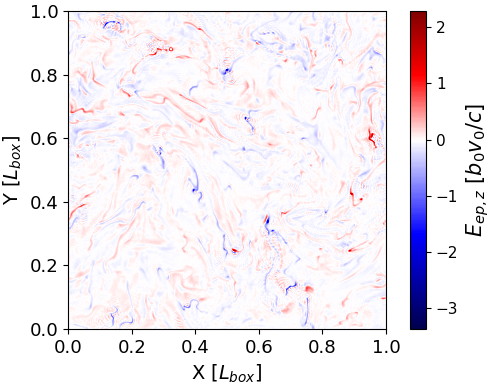}
        \caption{Cut in the $XY$ plane for a given $Z$ value.}
        \label{fetahorizontal}
    \end{subfigure}
    \caption{Cuts of the integration box at a given time showing the value of $E_{ep,z}$ with the color scale.}
    \label{fetas}
\end{figure*}

\begin{figure}
    \centering
    \includegraphics[scale=0.25]{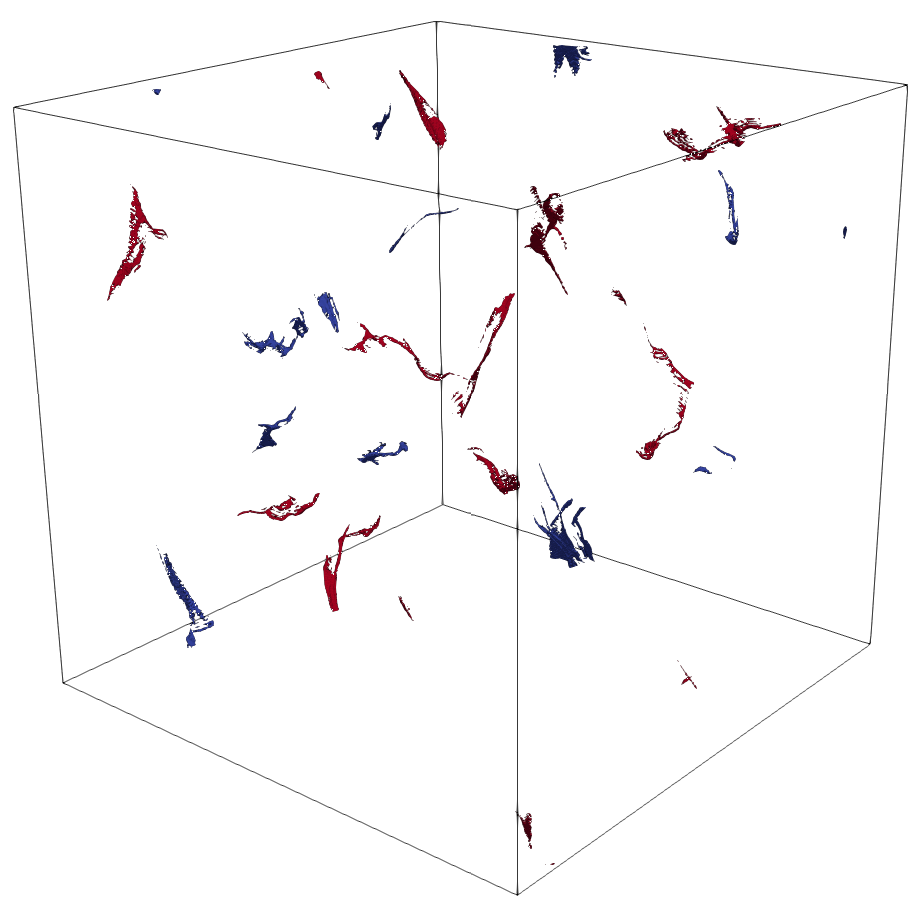}
    \caption{Plot of the integration box, where the colored structures are regions of strong positive (red) and negative (blue) values of $E_{ep,z}$, showing its filamentary structure.}
    \label{fig: fideitos}
\end{figure}

We can write the equation of motion for the particles in dimensionless form, using the MHD units defined in the previous subsection, we obtain:
\begin{gather}
    \frac{d\mathbf{v}}{dt} = \alpha(\mathbf{E}+\mathbf{v}\times\mathbf{B}) 
    \label{eq motion parts adim}
\end{gather}
where the dimensionless parameter $\alpha$ is:
\begin{gather}
    \alpha = Z\frac{m_p}{m}\frac{L}{d_{i}},
\end{gather}
with $Z$ the atomic number of the particle. The inverse of this parameter, $1/\alpha$, symbolizes the nominal gyroradius for particles with velocity $v_0$, measured in units of $L$. It also determines the range of scales present in the system, from the outer scale of turbulence to the particle gyroradius. If we consider the case where the particle is a proton, then $Z=1$ and $m=m_p$, which implies that $\alpha=L/d_{i}$. These are the particles that will be considered throughout this work. As for the value of $\epsilon=l_d/L$, it is reasonable to expect it to be quite small ($\epsilon\ll 1$), specially for space plasmas, and this can be a huge computational challenge. In the present work we have set $\epsilon\approx 1/60$.

The equation of motion of the particles (Eq. \eqref{eq motion parts adim}) is solved using a second-order Runge-Kutta method. To compute the value of the MHD fields in the position of the particles we use third-order spline interpolation. Initially the particles are distributed uniformly in the simulation box with a Gaussian velocity distribution function. The latter has a root mean square value of approximately $1.5v_0$. As for the simulation time, the particle's trajectories were integrated until $\langle \Delta z^2 \rangle^{1/2} = L_{box}$ to avoid periodicity effects. Here, $\langle\bullet\rangle$ is the average over all particles at fixed time. In every simulation performed, the MHD fields were evolved together with the particles but independently of them.

Figure \ref{averages_energies} depicts the evolution of the kinetic and magnetic energies of the fluid during the simulation time when particles are present. Figure \ref{fig:estacionarioSinH} shows this for the case where Hall effect is disregarded, and figure \ref{fig:estacionarioConH} illustrates the opposite. It is observed that the fluid is indeed in a statistically stationary state in both cases.

\section{\label{numerical results}Numerical Results}

In this section we focus on studying how the electric field terms, in particular the electronic pressure gradient and the Hall term, impact the dynamics of the test particles. To this end, we perform simulations with and without each of these terms (regarding the fluid itself, eqs \eqref{mhd1} to \eqref{mhd4}, and the evolution of the particles, eqs. \eqref{eq motion parts} and \eqref{ohm generalizado}).

\subsection{\label{relationship}Relationship between magnetic moment and kinetic energy}

We start by looking at the relationship between the magnetic moment $\mu$ and the energy $|\mathbf{v}|^2$ of the test particles. The magnetic moment is the ratio between the perpendicular energy of the particle and the background magnetic field:
\begin{equation}
    \mu = \frac{v_{\perp}^2}{B_0}.
\end{equation}
Here, $v_{\perp} = |v_x\hat{x}+v_y\hat{y}|$, meaning that we take the velocity perpendicular to the background field $B_0\hat{z}$. Considering the direction perpendicular to the whole magnetic field yields qualitatively the same results. At a fixed time, we can plot a two-dimensional histogram of both quantities for the $5\times 10^5$ particles. Displaying it for successive times, we are able to analyze the evolution of the particles distribution. In Figure \ref{fig:KvsMu_SinPE} this is shown for the case without the electronic pressure and Hall terms. From the definitions of $|\mathbf{v}|^2$ and $\mu$ we conclude that the particles falling on the straight line of slope $1/B_0 = 1/9$, marked with a dotted line in the figure, have exclusively perpendicular velocity. In the upper zone there cannot be any particle since it would imply $v_z^2<0$. On the other hand, the further the particles move away from this line, approaching towards the horizontal axis, the more relevant the velocity parallel to the background field becomes.

In figure \ref{fig:KvsMu_SinPE}, the vast majority of particles are very close to the line of slope $1/9$, i.e. their velocity is mainly perpendicular to $\mathbf{B}_0$. Also note that as time passes, particles increase both their magnetic moment and their energy, yet always remaining close to the line of slope $1/9$. This indicates a predominantly perpendicular energization (see \cite{dmitruk2004}).

Let us now consider the case in which we include the contribution of the electronic pressure to the electric field. In this case we still ignore the contribution of the Hall term. Repeating the same procedure as in the previous case, the histograms that we obtain are shown in figure \ref{fig:KvsMu_ConPE}.

Adding the effect of the electronic pressure on the particle dynamics shows a qualitative difference in the distributions. The particles are no longer clustered around the line of slope $1/9$, instead the distribution widens with respect to the previous case. It is thus evident that the electronic pressure causes an additional parallel acceleration on the particles that is not negligible. In fact, the population of particles on the horizontal line, corresponding to a purely parallel velocity, grows noticeably with respect to the case without electronic pressure. Nonetheless, considering that the color scale is logarithmic, we conclude that most of them still have a predominantly perpendicular velocity. Histograms corresponding to a simulation with the Hall term (not shown) are very similar to their non-Hall counterparts, showing its irrelevance in this respect. By definition, this term is proportional to $\mathbf{J}\times\mathbf{B}$, so it is perpendicular to the total magnetic field and, since fluctuations are small compared to $B_0$, the parallel component of the Hall term is negligible. The peculiarity of the electronic pressure is the unusual parallel acceleration it produces. Then, the Hall effect should not induce any behavior different from that of the figure \ref{fig:KvsMu_SinPE}. Indeed, when repeating the histograms for the case with Hall effect and without electronic pressure, what is found is qualitatively the same as in figure \ref{fig:KvsMu_SinPE}.

\begin{figure}
    \centering
    \begin{subfigure}[t]{0.47\textwidth}
        \centering
        \includegraphics[height=1.87in]{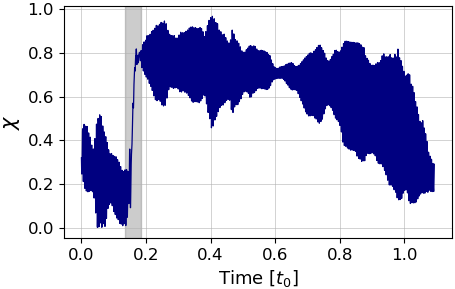}
        \caption{Value of $E_{ep,z}$ (parallel component of the electronic pressure contribution to the electric field) seen by an individual test particle throughout the simulation. The vertical gray area highlights a time period during which $E_{ep,z}$ experiences a peak.
        \vspace{4mm}}
        \label{fig:Fpez_partIndiv}
    \end{subfigure}\hfill
    ~ 
    \begin{subfigure}[t]{0.47\textwidth}
        \centering
        \includegraphics[height=2in]{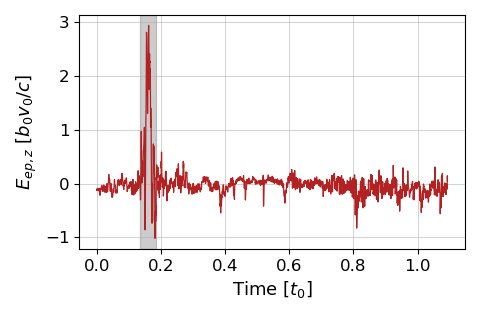}
        \caption{Value of $\chi$ (modulus of cosine pitch-angle) of an individual test particle throughout the simulation. The vertical gray area is the same as in figure \ref{fig:Fpez_partIndiv}. It also shows an abrupt increase in the value of $\chi$.}
        \label{fig:chi_partIndiv}
    \end{subfigure}
    \caption{Different magnitudes of the same individual test particle over the course of the simulation. The vertical gray area shows the same time period in both sub-figures.}
    \label{partIndiv}
\end{figure}

\subsection{\label{average}Average energies}

To better understand the impact of the electronic pressure, we analyze the final value of the average energies, that is, total energy $|\mathbf{v}|^2$, perpendicular $v_x^2 + v_y^2$ and  parallel $v_z^2$ energy, over all particles. This is shown in figure \ref{fig:comparacion energias finales} for all possible cases: Base case (without the Hall effect nor the electronic pressure term), only considering the Hall effect (H), only considering the electronic pressure term (EP), and taking both H and EP into account. Additionally, in figure \ref{fig:evolucion energias} the evolution of the average energies in the Base and the EP case is shown.

Observing figure \ref{fig:evolucion energias}, the first thing we note is the anisotropy between the perpendicular and the parallel components. The former grows significantly faster than the latter. On the other hand, it can be seen that a difference in the mean energies arises when the effect of the electronic pressure is incorporated. The total mean energy decrease slightly, the perpendicular energy also decreases, while the parallel energy has a noticeable growth.

From figure \ref{fig:comparacion energias finales} it is interesting to highlight that by adding only EP, the total energy (perpendicular+parallel) does not change dramatically. However, the parallel component grows at the expense of the perpendicular one. On the other hand, adding H generates a significant increase in the total energy. At first glance we could say that it impacts mostly on the perpendicular component. This is reasonable if we recall that the force it exerts is perpendicular to the total field ($\propto \textbf{J}\times\textbf{B}$). Although it has a component parallel to the guiding field $\mathbf{B}_0$, this is far smaller than the perpendicular one since $|\mathbf{b}|\ll\mathbf{B}_0$. However, the ratio between the two components is practically unchanged. Thus, the consequence seems to be a proportional increase of the energy. As for adding both H and EP, the effect that emerges is more complex. We note a reduction in the total energy and an increase in the relative importance of the parallel component that is larger than in all other cases; however, in what follows we will not go into further detail on the behavior derived from the interaction between EP and H.
This is because, although noticeable, their contribution seems secondary when studying mean values of test particles.
Clearly the bulk of the population is not much affected by these terms, in accordance to previous results\cite{gonzales2}.

In the remainder of this work, all the simulations shown are done considering both electronic pressure and Hall terms.

\subsection{\label{statistics}Statistics and structure of the electric field}

We know that electronic pressure generates a significant parallel acceleration on the test particles. The simplest hypothesis to explain this would be that the force generated by the electronic pressure was predominantly in the parallel direction. Let us call $\mathbf{E}_{ep}$ the contribution of EP to the electric field. That is,
\begin{gather}
    \mathbf{E}_{ep} = \frac{-\epsilon\mathbf{\nabla} p_e}{\rho\gamma M_S^2}.
\end{gather}
Thus, $\alpha\mathbf{E}_{ep}$ is the force generated by the electronic pressure. To check if the previous hypothesis is valid we can look at the distribution of the parallel and perpendicular components of $\mathbf{E}_{ep}$. In particular, we take $E_{ep,z}$ for the parallel part and $E_{ep,x}$ as representative of the perpendicular part. We could have chosen the y-component instead of x-component, since both are statistically equivalent. Figure \ref{fig:Fep para y perp} shows two probability density functions (PDFs), one for each component, of the field value over the position of all particles at all times.

Figure \ref{fig:Fep para y perp} refutes the assumption that $\mathbf{E}_{ep}$ might be predominant in the direction parallel to the background field. The standard deviation of the distribution in the perpendicular direction is  $\sigma^{ep}_x = 0.47\frac{b_0 v_0}{c}$, whereas its value in the parallel direction is $\sigma^{ep}_z = 0.15\frac{b_0 v_0}{c}$.  However, in figure \ref{fig:E' para y perp} we can see what happens if we repeat this analysis for the rest of the electric field, i.e., for
\begin{gather}
    \mathbf{E}' = \mathbf{E} - \mathbf{E}_{pe} = \mathbf{E} + \frac{\epsilon\nabla p_e}{\rho\gamma M_S^2} = \frac{\textbf{J}}{R_m} + \frac{\epsilon}{\rho}\textbf{J}\times\textbf{B} - \textbf{u}\times\textbf{B}.
\end{gather}
In such case, the deviations are $\sigma^{E'}_x = 16.42\frac{b_0 v_0}{c}$ and $\sigma^{E'}_z = 0.25\frac{b_0 v_0}{c}$. The ratio between them, $\sigma^{E'}_z/\sigma^{E'}_x = 0.015$, is much smaller than for $\mathbf{E}_{pe}$, where $\sigma^{ep}_z/\sigma^{ep}_x = 0.32$. We then conclude that, although the perpendicular component of $\mathbf{E}_{ep}$ is more significant than the parallel one, the latter is far more relevant compared to the rest of the electric field.

Therefore the electronic pressure generates a parallel acceleration in the particles because its contribution to the total field in the parallel direction is more significant than in the perpendicular direction. To check this, we compare the two figures above by plotting the PDFs of $E_{ep,x}/|E'_x|$ and $E_{ep,z}/|E'_z|$. In this way we can see the contribution of the electronic pressure compared to the rest of the electric field. This is shown in figure \ref{fig: Fpe sobre Eprima para y perp}. There we clearly observe that the dispersion of $E_{ep,z}/|E'_z|$ is larger than that of $E_{ep,x}/|E'_x|$. In fact, the deviations are $\sigma_z = 3.50\cdot 10^{5}$ and $\sigma_x = 4.61\cdot 10^{3}$ respectively.

Let us now study the structure of the field produced by the EP term. We are particularly interested in $E_{ep,z}$. In figure \ref{fetavertical} a cut in the $ZY$ plane of the integration box (at a fixed time) is shown. The color scale indicates the intensity of $E_{ep,z}$. Similarly, in figure \ref{fetahorizontal} a cut in the $XY$ plane is shown.

In figure \ref{fetavertical} note that $E_{ep,z}$ is composed of a near-zero, fairly monotonic background, within which appear scattered thin, elongated, wavy structures of significantly larger value. These are not flattened and stretched in the direction of the magnetic field, as would be the case with current sheets in plain MHD. Figure \ref{fig: fideitos} additionally shows a representation of them in 3D space. On the other hand, in the cut transverse to the background field $\mathbf{B}_0$ (figure \ref{fetahorizontal}) we see that the $\mathbf{E}_{ep}$ field has more structure than in the $ZY$ plane. This is to be expected in turbulent MHD flows under the effect of a guiding field. We also notice in this figure the presence of short scale fluctuations (wavy behavior) which we associate with the possible existence of fast magnetosonic waves in the fields (see for instance \cite{maia}).

In what follows we examine whether the particular structure of $E_{ep,z}$ corresponds to the dynamics of the individual test particles.

\begin{figure}
    \centering
    \includegraphics[scale=0.68]{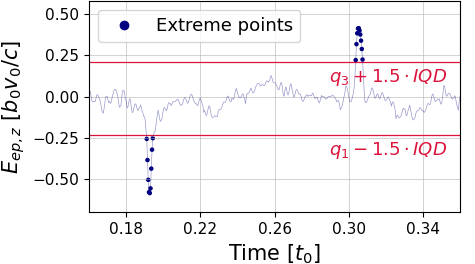}
    \caption{$E_{ep,z}$ field seen by a specific particle in a given time span. The horizontal lines represent the values of $(q_1 - 1.5 \cdot IQD)$ and $(q_3 + 1.5 \cdot IQD)$. Here, $q_{1,3}$ are the first and third quartiles of the set of $E_{ep,z}$ values seen by the particle over the entire simulation, and $IQD = q_3 - q_1$. Consecutive extreme points are also shown.}
    \label{fig: iqr}
\end{figure}

\begin{figure*}
    \centering
    \includegraphics[scale=0.65]{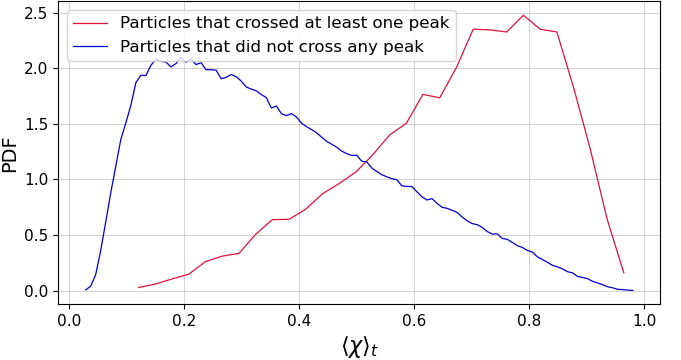}
    \caption{PDFs of the time-averaged value of $\chi$ (modulus of cosine pitch-angle), for the particles that crossed at least one peak of $E_{ep,z}$ (parallel component of electronic pressure contribution to the electric field) and for those particles that did not. Both histograms are normalized. The simulation was performed taking electronic pressure and Hall effect into account.}
    \label{fig:patadas pitch angle}
\end{figure*}

\subsection{\label{test}Test particle behaviour}

We now study the behavior of individual particles in order to deepen our understanding of the impact of the electronic pressure (EP). In particular, we are interested in the particles responsible for the change between the histograms in figures \ref{fig:KvsMu_SinPE} and \ref{fig:KvsMu_ConPE}. Thus, we focus on those particles whose ratio of magnetic momentum to kinetic energy deviates significantly from $1/B_0 = 1/9$ at least for some period of time. As an example, let us study a given particle that falls into this category. The value of $E_{ep,z}$ that this particles feels over time is shown in figure \ref{fig:Fpez_partIndiv}. And we can also see the evolution of the modulus of the cosine of its pitch-angle in figure \ref{fig:chi_partIndiv}. This quantity, which we will name $\chi$, is also defined as,
\begin{gather}
    \chi = \frac{|v_{\parallel}|}{|\mathbf{v}|}.
    \label{definicion chi}
\end{gather}
As we can see in the above definition, $\chi$ helps us measure the relative importance of both components of the velocity.

Figure \ref{fig:Fpez_partIndiv} depicts how the force due to EP in the vertical direction is fairly constant around zero, with a clear perturbation in the interval highlighted in gray. This perturbation is consistent with the crossing of one of the filamentary structures of $E_{ep,z}$ seen in figure \ref{fetavertical}. Both show similar magnitudes, around $2.5 b_0 v_0/c$. Moreover, it is expected for particles to experience an intense but brief force as they pass through one of these structures. This behavior occurs in a large portion of the particles whose ratio between $\mu$ and $|\mathbf{v}|^2$ departs from the value $1/9$.

Regarding figure \ref{fig:chi_partIndiv}, there is a clear boost of the particle's value of $\chi$ at about the same time when $E_{ep,z}$ is perturbed. Recalling equation \eqref{definicion chi}, this growth also indicates an increase in the importance of the parallel velocity over the perpendicular component. This seems to indicate that the electronic pressure generates a parallel acceleration in the particles through the thin structures of $E_{ep,z}$.

The goal now is to to test in a quantitative way whether there is a correlation between the particles that cross one of the thin structures of $E_{ep,z}$ and those whose average pitch-angle value is low. For that purpose we first need to discriminate between the particles that passed through at least one of the thin structures and those that did not. We focus on the $E_{ep,z}$ field seen by test particles throughout the entire simulation. For each one, we detect extreme values of $E_{ep,z}$ using the IQR method\cite{gelman1995bayesian}. With this approach, every point that is either smaller than $(q_1 - 1.5\cdot IQD)$ or greater than $(q_3 + 1.5\cdot IQD)$ is considered to be extreme, as illustrated in figure \ref{fig: iqr} for a specific particle. Here, $q_1$ and $q_3$ are the first and third quartiles of the set of $E_{ep,z}$ values seen by the specific particle, and $IQD$ is the inter-quartile distance \cite{stats_book}. Then, groups of consecutive extreme points are identified, two of which are shown in figure \ref{fig: iqr} as blue dots. Every one of these can potentially be one of the thin structures of the EP field. To determine whether they are or not, we calculate the energy supplied by the $E_{ep,z}$ field for all points in each group. That is, 
\[\int \mathbf{E}_{ep}\cdot \mathbf{v} dt \approx \sum E_{ep,z}v_z\Delta t\]
where the sum goes over all extreme points within a group. We consider a given group to be a peak of $E_{ep,z}$ (i.e., one of the thin structures) if its delivered energy is greater than a given threshold.

To define this threshold, we consider the fact that the energy supplied by crossing one of these structures could also be written as $\mathcal{E}_{ep, z}\Delta z$ where $\mathcal{E}_{ep, z}$ is a characteristic value of $E_{ep, z}$ in the structure and $\Delta z$ its characteristic length in the z-direction.
Using a similar IQD criterion, we identify extreme values of $E_{ep,z}$, thus allowing us to discriminate structures in the underlying 3D field.
We then extract only the extreme values and set $\mathcal{E}_{ep, z}$ to their rms value.
For the lenght $\Delta z$, we define a new field $E_{ep,z}'$ containing only the extreme values (and $0$ otherwise) and calculate its autocorrelation function $C(\delta_z) = \langle E_{ep, z}'(\mathbf{x}+\delta_z \hat{z}, t)E_{ep, z}'(\mathbf{x}, t) \rangle_{\mathbf{x}, t}$.
We then define $\Delta z$ as the value of its first zero-crossing, which yields $\Delta z \approx L_{box}/26$ and is compatible with the structures observed in figure \ref{fetas}, specially when taking into account their variable orientation.
The last observation is relevant given that most particles do not cross the filaments through their thinnest direction, which is clearly smaller than $L_{box}/26$.
We verify this threshold by analyzing some individual particle trajectories and find it appropriate.

Since we are trying to correlate the thin structures of the EP field with high values of $\chi$, we are not interested in those particles that have a big $\chi$ at the beginning of the simulation. These will have a value close to one not due to the EP term but because of their initial condition. Regarding this, all particles that initially have a parallel energy $v_{\parallel}^2$ greater than the perpendicular one $v_{\perp}^2$, will be excluded. In other words, we keep only those that satisfy $\chi\leq 1/\sqrt{2}$ at the beginning. After doing this, the $500,000$ particles have reduced to a number of $360,281$.

From the total of $360,281$ particles that we kept, $10,200$ were found to have passed through at least one thin structure of $E_{ep,z}$. For the two groups of particles (the ones that crossed a peak and those that did not) we plot a PDF of the average over time of $\chi$ in figure \ref{fig:patadas pitch angle}. Let us point out that if this quantity equals $1$ then the particle's velocity is parallel to $B_0$, and the contrary happens if it equals $0$. We choose the cosine of the pitch-angle instead of simply the pitch-angle because the former is uniformly distributed for an isotropic 3D velocity distribution.

\begin{figure}
    \centering
    \begin{subfigure}[t]{0.47\textwidth}
        \centering
        \includegraphics[height=1.83in]{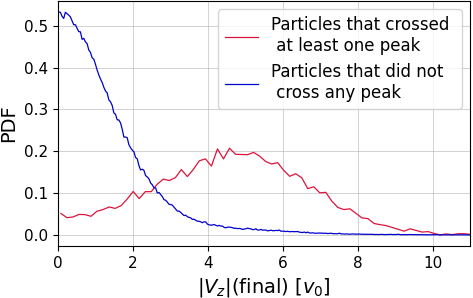}
        \caption{PDF of the final value of $|V_z|$ (parallel velocity of the particle), distinguishing between particles that went through at least one thin structure of the EP field and those that didn't.
        \vspace{4mm}}
        \label{fig:Vz finales patada y no}
    \end{subfigure}\hfill
    ~ 
    \begin{subfigure}[t]{0.45\textwidth}
        \centering
        \includegraphics[height=1.83in]{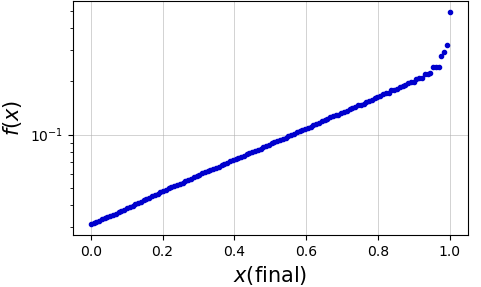}
        \caption{Probability $f(x)$ of a particle having gone through at least one thin structure of the EP field given that its final value of $\chi$ (modulus of cosine pitch-angle) satisfies $\chi\geq x$.}
        \label{fig:f(x)}
    \end{subfigure}
    \caption{Overview of the effect of electronic pressure on test particles.}
    \label{patadas y no patadas}
\end{figure}

There is a clear separation of both PDFs in figure \ref{fig:patadas pitch angle}. Particles that did not go through a peak of $E_{ep,z}$ cluster towards lower values of $\langle\chi\rangle_t$, while the opposite is true for the particles that did cross at least one peak. It is then evident that there is a correlation between the thin structures of the EP field and the parallel acceleration of particles due to the presence of EP. This seems to confirm that this type of structure is the primary method by which electronic pressure pushes particles in the direction of the background magnetic field. Particles going through them experience a strong but quick force in the $\hat{z}$ direction, changing their pitch-angle.

We can also examine how the probability distribution of the final parallel velocity changes from one set of particles to the other. This is shown in figure \ref{fig:Vz finales patada y no} for the absolute value of $V_z$. In the case of particles that did not cross any thin structure, it is reasonable to expect for their distribution to peak at the origin since they did not experience any extraordinary force in the $\hat{z}$ direction. The distribution of the others, on the contrary, should be centered on a higher velocity value. Indeed, both of these things are reflected in figure \ref{fig:Vz finales patada y no}, showing once again the effect of electronic pressure in parallel acceleration.

To further quantify the relevance of EP, let us define the function $f(x)$ as the conditional probability of a particle having crossed a peak given that its final value of $\chi$ satisfies $\chi \geq x$. To find it we first select particles having $\chi\geq x$ at the end of the simulation, and then we determine the fraction that crossed a peak. Whether a particle went through a peak or not is established using the previously explained method.

In figure \ref{fig:f(x)} the function $f(x)$ is plotted for $0\leq x\leq 1$. We first notice that, as expected, it is an increasing function. This indicates that greater values of the final $\chi$ correspond to a higher likelihood of the particle having crossed a peak of the EP field. In fact, half of high pitch-angle particles have interacted with these structures. In addition, the linearity in the semi-log graph shows an exponential behaviour of $f(x)$ up to $x \sim 0.9$, with an exponent of $2.07\pm 0.02$. %El 0.01 es un desvío estandar del error (que sale del ajuste).

\section{\label{conclusiones}Conclusions}

In this paper we used direct numerical simulations to investigate the effect that electronic pressure (EP) has on the acceleration of test particles immersed in a turbulent MHD field under the influence of a strong background magnetic field $\mathbf{B}_0$. We found that EP generates an acceleration parallel to $\mathbf{B}_0$ in a portion of the particles, that is unusual in the most common case where the electronic pressure term and the Hall term are neglected. In particular, it was found that considering the EP term increases the percentage of the particles' average parallel energy, to the detriment of the perpendicular component.
This effect is secondary when considering mean energies and affects only a small fraction of the population.

Additionally, we studied the statistics of the electric field felt by the particles in every time step of the simulation. We found that the contribution of the electronic pressure $\mathbf{E}_{ep}$ to the electric field is \textit{not} predominant in the parallel direction, as might have been expected, but that its contribution to the total electric field is more significant in the parallel direction than in the perpendicular one. This helps us explain the above results. As for the structure of the $E_{ep,z}$ field, it was found that, in the planes parallel to $\textbf{B}_0$, it is composed of a monotonous and low-value background. Yet, on top of it we found thin and elongated structures whose absolute value is much higher, which we suppose could be related to fast magnetosonic wave interactions\cite{maia}.

Studying the behavior of individual test particles we noticed that those particles undergoing an unusual parallel acceleration also appeared to have crossed one of the thin structures of $E_{pe,z}$. We separated the ones that went through one of this structures from those that did not using a criterion based on the supplied energy of the EP field. Then, looking at the average pitch-angle's cosine for the two groups of particles, a clear separation between both distributions was noted. While those particles that crossed a thin structure were prone to have a large value of $\langle\chi\rangle_t$ (that is, a predominance of the parallel component of the velocity), the opposite was true for the particles that did not. This is evidence of the parallel acceleration method of the $E_{ep,z}$ field; a strong and brief thrust to those particles that pass through its thin and elongated structures located on top of the monotonic and near-zero background. Since these structures are thin and scattered in space, it is natural to conceive particles that manage to avoid them, and thus do not suffer any extra parallel push.

In general, we have been able to measure the impact that the electronic pressure term in the generalized Ohm's law has on test particles. Although this effect is often neglected, we now have a better idea of what types of consequences are being overlooked and what their extent is. We believe that this contributes to the understanding of the behavior of charged particles in space plasmas, in particular, the theoretical results presented here could be useful for the interpretation of recent observational results relating to proton beams along the mean magnetic field\cite{carlos1,carlos2,carlos3,pecora}.

\nocite{*}
\bibliography{aipsamp}% Produces the bibliography via BibTeX.

\end{document}